\documentclass[11pt]{article}
\usepackage{moriond,epsfig}

\usepackage{rotating}
\usepackage{caption}

\def\ifb{\ensuremath{ \rm{fb}^{-1}}}

\newcommand{\Eslash}{\mbox{$E \kern-0.6em\slash$                }}
\newcommand{\etmiss}{\mbox{\ensuremath{\Eslash_{\kern-0.3emT}\!} }}
\newcommand{\dzero}{\mbox{D\O}}
\newcommand{\GeV} {\ensuremath{\mathrm{Ge\kern -0.1em V}}}

\def\be{\begin{equation}}
\def\ee{\end{equation}}
\def\bea{\begin{eqnarray}}
\def\eea{\end{eqnarray}}

\newcommand{\met}       {\mbox{$\not\!\!E_T$}}

\newcommand{\ppbar}     {\mbox{$p\bar{p}$}}
\newcommand{\ttbar}     {\mbox{$t\bar{t}$}}

\newcommand{\comphep}   {\sc comphep}
\newcommand{\singletop} {\sc singletop}
\newcommand{\pythia}    {\sc pythia}
\newcommand{\alpgen}    {\sc alpgen}

\begin{document}

\begin{table}
\begin{flushleft}
\begin{minipage}{5.5 in}
\begin{tabular}{lr}
{\rightline{\hspace{-0.5in}{FERMILAB-CONF-12-241-E}}} \\
{\rightline{\hspace{-0.5in}{{\dzero} NOTE 6333}}} \\
{\rightline{\hspace{-0.5in}{May 2012}}} \\ 
\end{tabular}
\end{minipage}
\end{flushleft}
\end{table}

\vspace*{2cm} 
\title{Single Top Quark Production at \dzero}

\author{ Jyoti Joshi (for the \dzero~collaboration)} 
\address{ University of California, Riverside, CA 92521-0413, USA}

\maketitle\abstracts{ We present new measurements of the single top quark production cross section in \ppbar~collisions at $\sqrt{s} = 1.96~TeV$ using data corresponding to 5.4 \ifb of integrated luminosity collected by the {\dzero} detector at the Fermilab Tevatron collider. The large mass of the top quark, close to the electroweak symmetry-breaking scale, makes it a good candidate for probing physics beyond the Standard Model, including possible anomalous couplings. We examine the data to study the Lorentz structure of the $Wtb$ coupling, and find that the data prefer the left-handed vector coupling, and set upper limits on the anomalous couplings.}


\section{Introduction}\label{sec:intro}
At hadron colliders, top quarks are produced as $\ttbar$ pairs via the strong interaction or singly via the electroweak interaction~\cite{ttbar-xsec,singletop-xsec-kidonakis}. Electroweak single top quark production was observed by the {\dzero} and CDF collaborations~\cite{stop-obs-2009} in 2009. Electroweak production of top quarks at the Tevatron proceeds mainly via the decay of a time-like virtual $W$ boson 
accompanied by a bottom quark in the $s$-channel ($tb = t\bar{b}+\bar{t}b$)~\cite{singletop-cortese}, or via the exchange of a space-like virtual $W$ boson between a light quark and a bottom quark in the $t$-channel ($tqb = tq\bar{b}+\bar{t}qb$, where $q$ refers to the light quark)~\cite{singletop-willenbrock-yuan}. 
A third process, usually called ``associated production,'' in which the top quark is produced together with a $W$ boson, has a negligible cross section at the Tevatron~\cite{singletop-xsec-kidonakis} and is therefore not considered in this analysis.
For a top quark mass of $172.5$~GeV, the Standard Model (SM) prediction of the single top production rates at  next-to-leading order with soft-gluon contributions at next-to-next-to-leading order are $1.04 \pm 0.04$~pb ($s$-channel) and $2.26 \pm 0.12$~pb ($t$-channel)~\cite{singletop-xsec-kidonakis}. Single top quark production is distinct from $\ttbar$ pair production since it comes from an
electroweak $Wtb$ vertex instead of a strong $gtt$ vertex and hence it provides a unique probe to study the interactions of the top quark with the $W$ boson. 


\section{Event Selection}\label{sec:selection}

The results presented here uses 5.4 \ifb of data collected with {\dzero} detector between 2002 to 2009. 
The single top quark events are expected to contain at least one $b$ quark jet from the decay of the top quark and a second $b$ quark jet in the $s$-channel, or a light quark jet and a spectator $b$ quark jet for the $t$-channel. In both cases, gluon radiation can give rise to additional jets. 
Events are selected that contain one jet with transverse momentum $p_T>25\;\rm GeV$ and at least a second jet with $p_T>15\;\rm GeV$, both within pseudorapidity $|\eta|<3.4$. Events are also required to contain exactly one isolated high-$p_T$ electron or muon that originates from the $\ppbar$ interaction vertex and satisfies the following acceptance criteria: for the electron $|\eta| < 1.1$ and $p_T > 15 (20)\;\rm GeV$ for events with 2 (3 or 4) jets; for the muon $|\eta| < 2.0$ and $p_T > 15\;\rm GeV$.
The $\met$ is required to be in the range of $(20,200)\;\rm GeV$ for events with 2 jets and
$(25,200)\;\rm GeV$ for events with 3 or 4 jets. 
The SM predicts a purely left-handed vector coupling ($f_{L_{V}}$) at the $Wtb$ vertex, while the most general, lowest dimension Lagrangian~\cite{lagrangian} allows right-handed vector ($f_{R_{V}}$) and left-handed tensor ($f_{L_{T}}$) or right-handed tensor ($f_{R_{T}}$) couplings as well.
Single top quark signal events with the SM and anomalous $Wtb$ couplings are modeled using the {\comphep}-based MC event generator {\singletop}~\cite{singletop-mcgen}. The anomalous $Wtb$ couplings are taken into account in both production and decay in the generated samples. 
The theoretical cross sections for anomalous single top quark production
($s$+$t$-channel) with $|V_{tb}| \simeq 1$ are $3.1\pm0.3$~pb if $f_{R_V}=1$, $9.4\pm1.4$~pb if $f_{L_T}=1$ or $f_{R_T}=1$, and $10.6\pm0.8$~pb if $f_{L_T}=f_{L_V}=1$~\cite{dudko-boos}, all other 
couplings are set to zero when calculating these cross sections. 
The {\ttbar}, $W$+jets, and $Z$+jets backgrounds are simulated using
the {\alpgen} leading-log MC event generator~\cite{alpgen}, with
{\pythia}~\cite{pythia} used to model hadronization. 
The main contributions to the systematic uncertainty on the predicted number of events arise from the signal modeling, the jet energy scale (JES), jet energy resolution (JER), corrections to $b$-tagging efficiency and the correction for jet-flavor composition in $W$+jets events. 
The total systematic uncertainty on the background is 11\%. Table~\ref{tab:event-yields} lists the numbers of events expected and observed for each process as a function of jet multiplicity.

\begin{table}[!h!tbp]
\caption{Numbers of expected and observed events in $5.4\;\rm fb^{-1}$ of integrated luminosity, with uncertainties including both statistical and systematic components. The single top quark contributions are
normalized to their theoretical predictions.
}
\label{tab:event-yields}
\begin{center}
\begin{tabular}{lrclrclrcl}
\hline
Source     & \multicolumn{3}{c}{2 jets} & \multicolumn{3}{c}{3 jets} & \multicolumn{3}{c}{4 jets} \\
\hline
$tb\ $  ($f_{L_T}=1$)      & 730 &$\pm$& 38 & 316 &$\pm$& 25 & 92 &$\pm$& 14 \\
$tqb$   ($f_{L_T}=1$)      & 117 &$\pm$& 6.2 & 86 &$\pm$& 8.6 & 40 &$\pm$& 5.8 \\
$tb\ $  ($f_{L_V}=f_{L_T}=1$)      & 607 &$\pm$& 31 & 284 &$\pm$& 21 & 86 &$\pm$& 13 \\
$tqb$   ($f_{L_V}=f_{L_T}=1$)     & 268 &$\pm$& 15 & 167 &$\pm$& 16 & 67 &$\pm$& 10 \\
$tb\ $  ($f_{R_V}=1$)        & 105 &$\pm$& 6.0 & 43 &$\pm$& 3.8 & 12 &$\pm$& 1.9 \\
$tqb$   ($f_{R_V}=1$)       & 122 &$\pm$& 7.2 & 61 &$\pm$& 5.3 & 22 &$\pm$& 3.7 \\
$tb\ $  ($f_{R_T}=1$)       & 756 &$\pm$& 42 & 344 &$\pm$& 27 & 103 &$\pm$& 15 \\
$tqb$   ($f_{R_T}=1$)       & 103 &$\pm$& 5.8 & 67 &$\pm$& 6.3 & 28 &$\pm$& 4.4 \\
\hline
$tb\ $  (SM, $f_{L_V}=1$)       & 104 &$\pm$& 16 & 44 &$\pm$& 7.8 & 13 &$\pm$& 3.5 \\
$tqb$   (SM, $f_{L_V}=1$)      & 140 &$\pm$& 13 & 72 &$\pm$& 9.4 & 26 &$\pm$& 6.4 \\
${\ttbar}$ & 433 &$\pm$& 87 & 830 &$\pm$& 133 & 860 &$\pm$& 163 \\
$W$+jets   & 3,560 &$\pm$& 354 & 1,099 &$\pm$& 169 & 284 &$\pm$& 76 \\
$Z$+jets and dibosons & 400 &$\pm$& 55 & 142 &$\pm$& 41 & 35 &$\pm$& 18 \\
Multijets  & 277 &$\pm$& 34 & 130 &$\pm$& 17 & 43 &$\pm$& 5.2 \\
\hline
Total SM prediction & 4,914 &$\pm$& 558 & 2,317 &$\pm$& 377 & 1,261 &$\pm$& 272 \\
\hline
Data       & \multicolumn{3}{c}{4,881} & \multicolumn{3}{c}{2,307} & \multicolumn{3}{c}{1,283} \\
\hline
\end{tabular}
\end{center}
\end{table}

\section{Single Top Quark Production Cross Section Measurement}\label{sec:xsec}

Since the expected single top quark contribution is smaller than the uncertainty on the background count prediction, multivariate analysis (MVA) methods are used to improve the discrimination between signal and background events. Three different MVA techniques are used for the cross section extraction: (i) boosted decision trees (BDT)~\cite{decision-trees}, (ii) bayesian neural networks (BNN)~\cite{bayesianNNs}, and (iii) neuroevolution of augmented topologies (NEAT)~\cite{neat}. All the three methods use the same data and background model considering the same sources of systematic uncertainties. Each MVA method is trained separately for the two single top quark production channels: for the $tb$~($tqb$) discriminants, with $tb$~($tqb$) considered signal and $tqb$~($tb$) treated as a part of the background. To achieve the maximum sensitivity, the three methods are combined to construct a new discriminant using a seccond BNN for each channel, for $tb$, $tqb$, and $tb+tqb$ events. The combined $tb+tqb$ discriminant is constructed by taking input from the six discriminant outputs of BDT, BNN and NEAT that are trained separately for the $tb$ and $tqb$ signal. The single top quark production cross section is measured using a Bayesian inference approach~\cite{singletop-prd-new,singletop-plb-new}. To measure the individual $tb$ ($tqb$) production cross section, a one-dimensional (1D) posterior probability density function is constructed with the $tqb$ ($tb$) contribution normalized with Gaussian priors centered in the predicted SM cross section for each individual MVA method and also for their combination. To measure the total single top quark production cross section of $tb$+$tqb$, a 1D posterior probability density function is constructed assuming the production ratio of $tb$ and $tqb$ predicted by the SM. Fig.~\ref{fig:posterior1d} shows the resulting expected and observed posterior density distributions for $tb$, $tqb$ and $tb+tqb$ for the combined discriminants.

\begin{figure*}[!ht]
\includegraphics[width=0.30\textwidth]{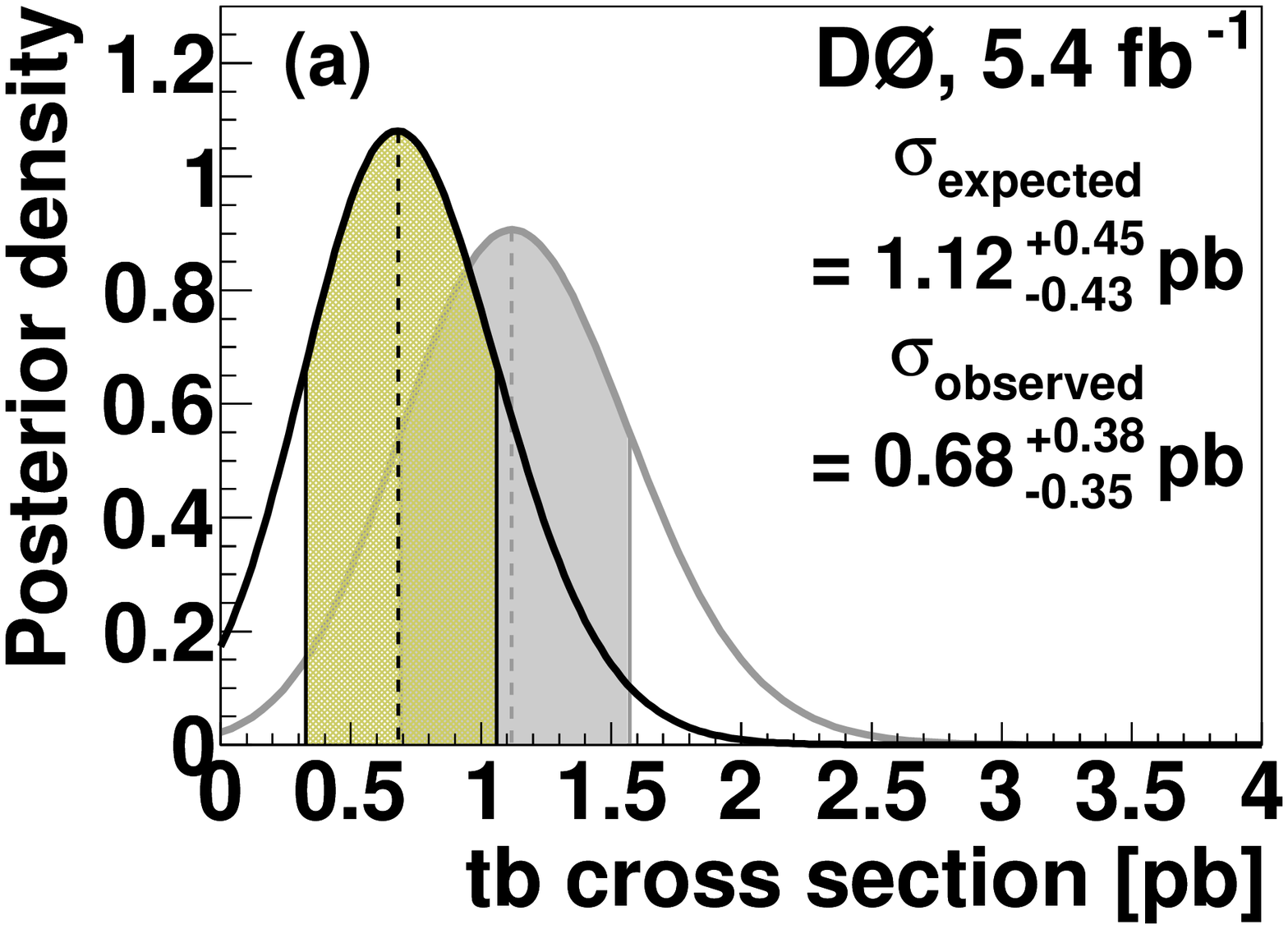}
\includegraphics[width=0.30\textwidth]{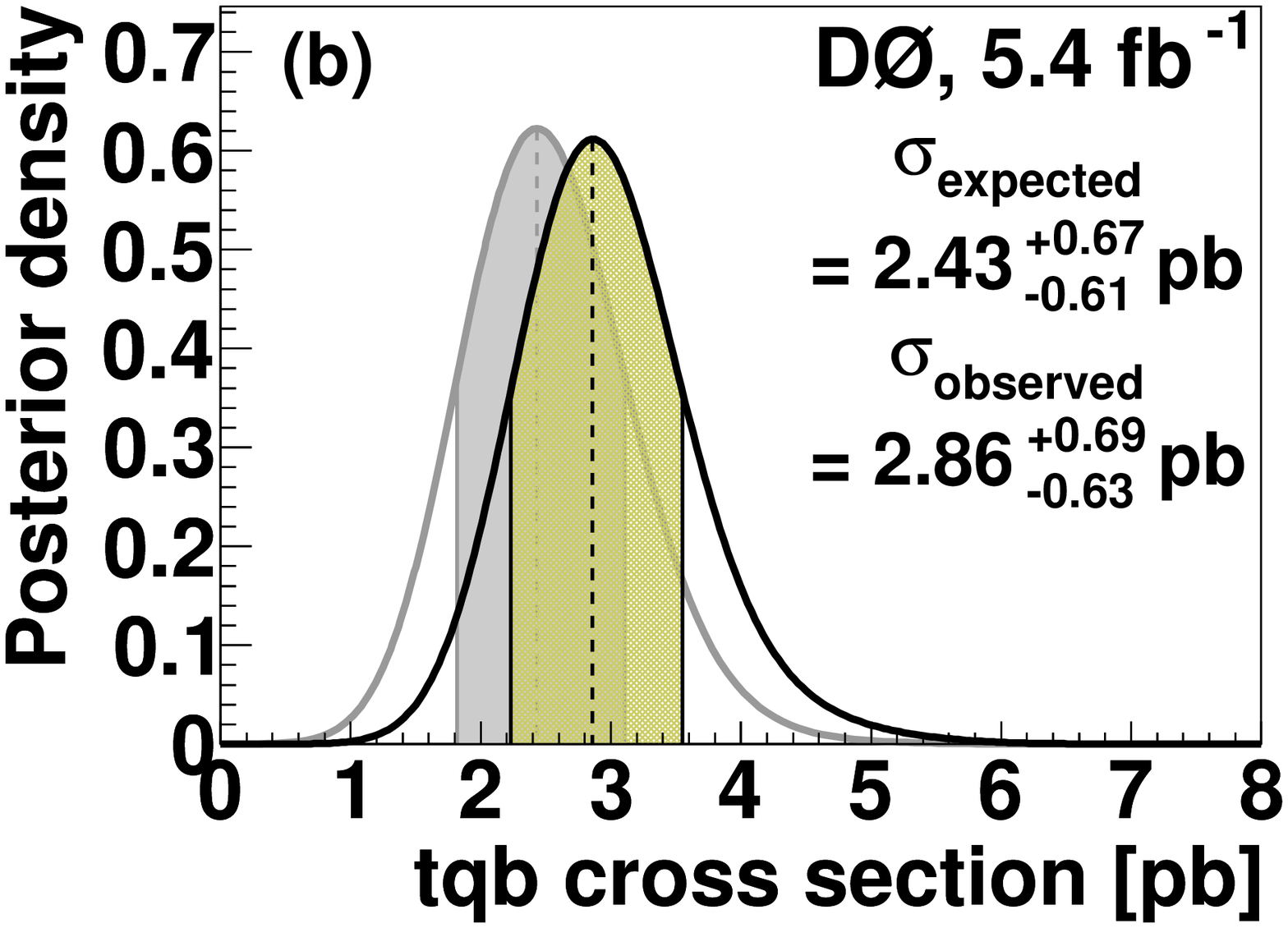}
\includegraphics[width=0.30\textwidth]{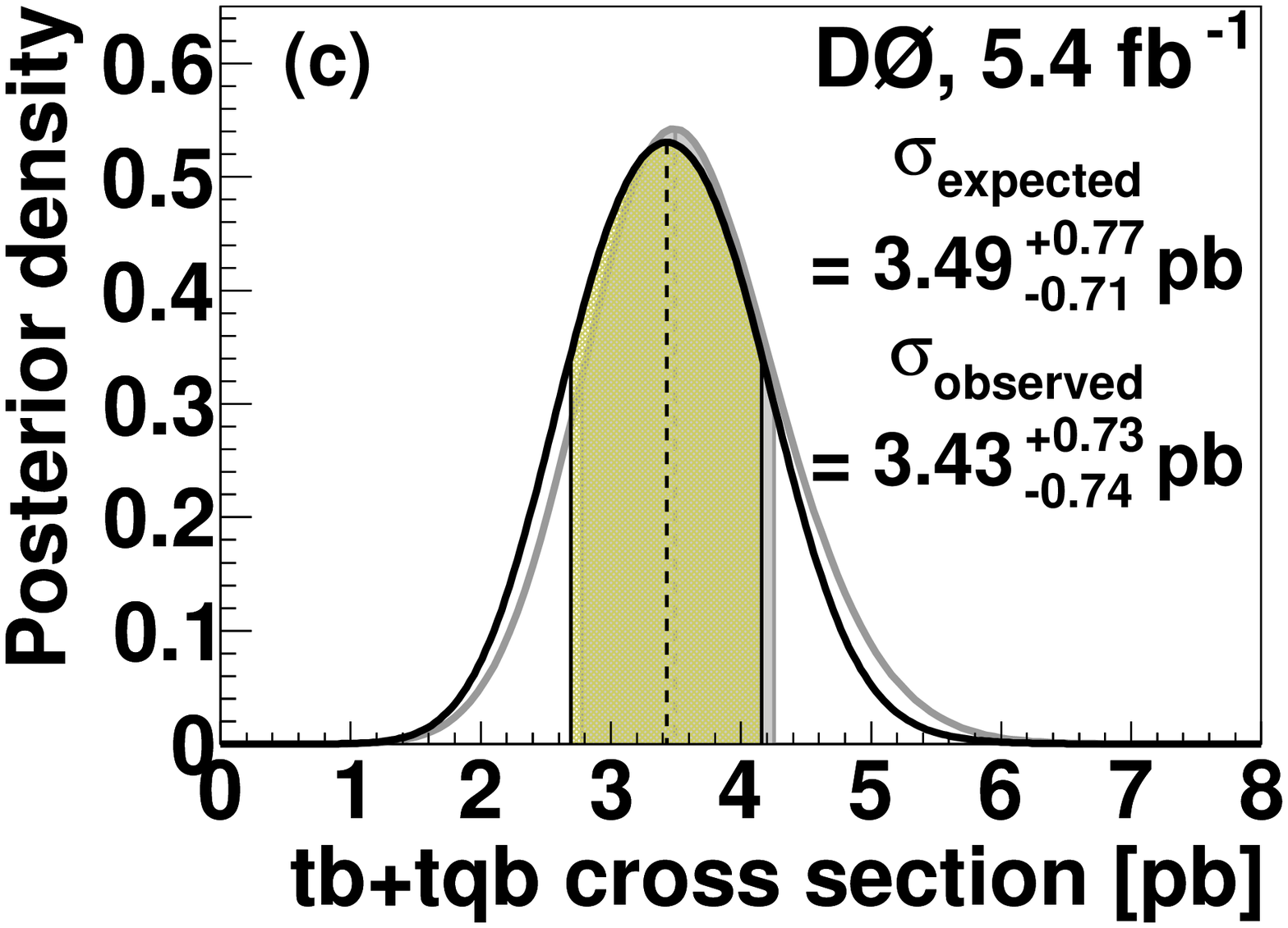}
\vspace{-0.1in}
\caption{The expected (grey) and observed (black) posterior probability densities
for (a) $tb$, (b) $tqb$, and (c) $tb$+$tqb$ production.
The shaded bands indicate the 68\% C.L.s from the peak values.}
\label{fig:posterior1d}
\end{figure*}

\section{Anomalous $Wtb$ Couplings}\label{sec:wtb}

As discussed earlier, by studying single top quark production, we can test whether the $Wtb$ coupling is pure left-handed vector in form, or whether there are right-handed vector, or left- or right-handed tensor components present. Assuming single top quarks are produced only via $W$ boson exchange, the single top quark cross section is directly proportional to the square of the effective $Wtb$ coupling. Moreover, the event kinematics and angular distributions are also sensitive to the existence of anomalous top quark couplings~\cite{dudko-boos}. Therefore, direct constraints on anomalous couplings can be obtained by measuring single top quark production. An analysis has been performed using the same dataset, event selection and background as the cross section measurement analysis, between SM background (including SM single top quark) and anomalous single top quark production as a signal, to set limits on the $Wtb$ coupling for other than a pure left-handed vector form~\cite{singletop-wtb-new}. A BNN is used to discriminate between signal and background. A Bayesian statistical approach is followed to compare data to the signal predictions given by different anomalous couplings. A two-dimensional (2D) posterior probability density
is computed as a function of  $|V_{tb} \cdot f_{L_V}|^2$ and $|V_{tb} \cdot f_{X}|^2$, where $f_{X}$ is any of the three nonstandard couplings and $V_{tb}$ is the Cabibbo-Kobayashi-Maskawa matrix element~\cite{bib:CKM}. The two dimensional (2D) limit contours are shown in Fig.~\ref{fig:measfullsys_2D}. 
We measure upper limits $|V_{tb} \cdot f_{L_T}|^2<0.06$, $|V_{tb} \cdot f_{R_V}|^2<0.93$ and
$|V_{tb} \cdot f_{R_T}|^2<0.13$ at 95\% C.L. after integrating the 2D posterior over $|V_{tb} \cdot f_{L_V}|^2$.

\begin{figure*}[!h!tbp]
\includegraphics[width=0.30\textwidth]{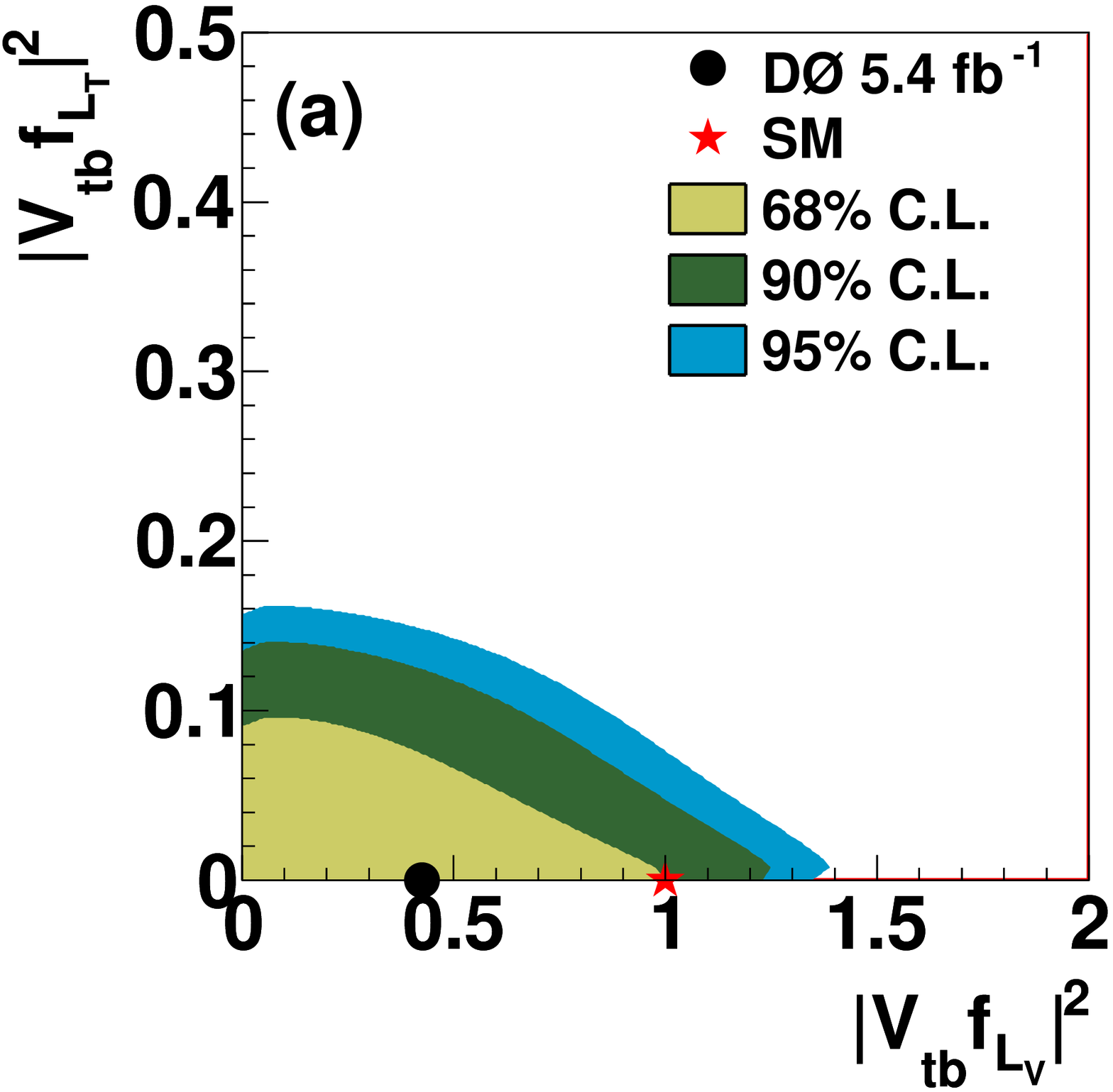}
\includegraphics[width=0.30\textwidth]{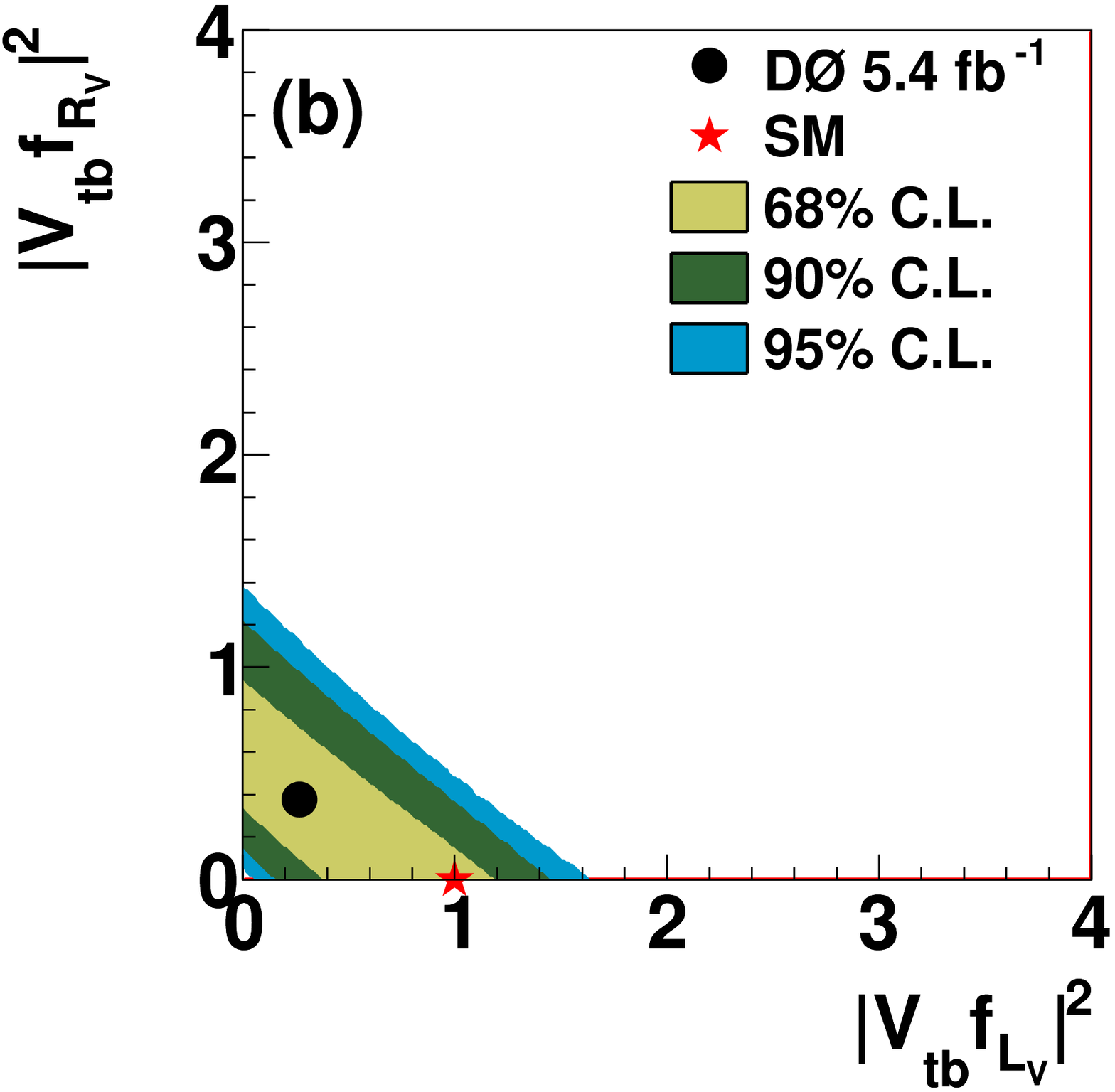}
\includegraphics[width=0.30\textwidth]{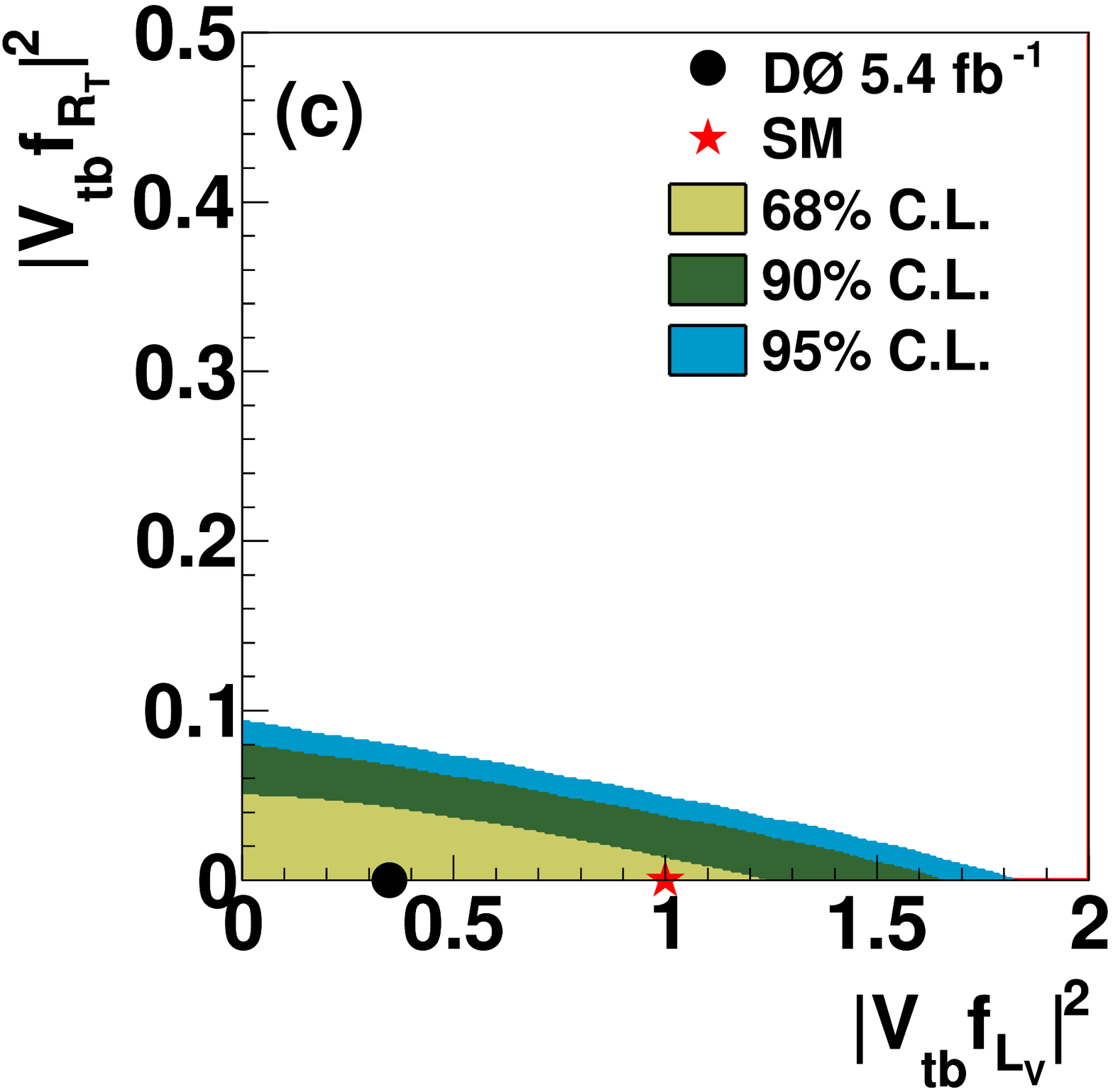}
\caption{Two-dimensional posterior probability density distributions for the anomalous couplings. (a) ($L_V$,$L_T$) scenario, (b) ($L_V$,$R_V$) scenario, and (c) ($L_V$,$R_T$) scenario.} 
\label{fig:measfullsys_2D}
\vspace{-0.3in}
\end{figure*}

\section{Summary}\label{sec:result}

In summary, we have measured the single top quark production cross section using 5.4~fb$^{-1}$ of data collected by
the {\dzero} experiment at the Fermilab Tevatron Collider. 
For $m_t=172.5\;\rm GeV$, we measure the cross sections for $tb$ and $tqb$ production
to be $0.68^{+0.38}_{-0.35}\;\rm pb$ and $2.86^{+0.69}_{-0.63}\;\rm pb$
assuming, respectively, $tqb$ and $tb$ production rates as predicted
by the SM. The total $tb+tqb$ cross section, assuming the SM ratio between $tb$ and $tqb$ production is $3.43^{+0.73}_{-0.74}\;\rm pb$.
Also, we searched for the anomalous $Wtb$ couplings in single top quark production and found 
no evidence for them and set 95\% C.L. limits on these couplings. 
This result represents the most stringent direct constraints on anomalous $Wtb$ interactions.


\section*{References}
\begin{thebibliography}{99}
\bibitem{ttbar-xsec}
S.~Moch and P.~Uwer,
Phys.\ Rev.\ D {\bf 78}, 034003 (2008). 

\bibitem{singletop-xsec-kidonakis}
N.~Kidonakis,
Phys.\ Rev.\ D {\bf 74}, 114012 (2006). 

\bibitem{stop-obs-2009}
V.~M.~Abazov {\it et al.} ({\dzero}~Collaboration),
Phys.\ Rev.\ Lett.\ {\bf 103}, 092001 (2009); T.~Aaltonen {\it et al.} (CDF Collaboration),
Phys.\ Rev.\ Lett.\ {\bf 103}, 092002 (2009).

\bibitem{singletop-cortese}
S.~Cortese and R.~Petronzio,
Phys.\ Lett.\ B {\bf 253}, 494 (1991).

\bibitem{singletop-willenbrock-yuan}
C.-P.~Yuan,
Phys.\ Rev.\ D {\bf 41}, 42 (1990).

\bibitem{lagrangian}
G.~L.~Kane, G.~A.~Ladinsky, and C.-P. Yuan,  Phys. Rev. D {\bf 45}, 124 (1992);
J.~A.~Aguilar-Saavedra, Nucl.\ Phys.\ B\ {\bf 812}, 181 (2009).

\bibitem{singletop-mcgen}
E.~E.~Boos {\it et al.},
Phys.\ Atom.\ Nucl.\ {\bf 69}, 1317 (2006). We used {\singletop}
version 4.2p1.

\bibitem{dudko-boos}
E.~Boos, L.~Dudko, and T.~Ohl,
Eur.\ Phys.\ J.\ C {\bf 11}, 473 (1999).

\bibitem{alpgen}
M.~L.~Mangano {\it et al.},
J.\ High Energy Phys.\ {\bf 07}, 001 (2003). 

\bibitem{pythia}
T.~Sj\"{o}strand, S.~Mrenna, and P.~Skands,
J.\ High Energy Phys.\ {\bf 05}, 026 (2006). 

\bibitem{decision-trees}
L.~Breiman {\it et al.},
{\it Classification and Regression Trees}
(Wadsworth, Stamford, 1984).

\bibitem{bayesianNNs}
R.M.~Neal,
{\it Bayesian Learning for Neural Networks}
(Springer-Verlag, New York, 1996).

\bibitem{neat}
K. O. Stanley \& R. Miikkulainen, 
Evolutionary Computation {\bf 10}, 99 (2002).

\bibitem{singletop-prd-new}
V.~M.~Abazov {\it et al.} ({\dzero}~Collaboration), Phys. \ Rev. \ D {\bf 84}, 112001 (2011).

\bibitem{singletop-plb-new}
V.~M.~Abazov {\it et al.} ({\dzero}~Collaboration), Phys. \ Lett. \ B {\bf 705}, 313 (2011). 

\bibitem{singletop-wtb-new}
V.~M.~Abazov {\it et al.} ({\dzero}~Collaboration), Phys. \ Lett. \ B {\bf 708}, 21 (2011).

\bibitem{bib:CKM}
N.~Cabibbo, Phys.\ Rev.\ Lett.\ {\bf 10}, 531 (1963); M.~Kobayashi and T.~Maskawa, Prog.\ Theor.\ Phys.\ {\bf 49}, 652 (1973).

\end{thebibliography}
\end{document}